# Self-organising Dissipative Polymer Structures

Andrei N. Yakunin
Laboratory of Polymer Materials,
Karpov Institute of Physical Chemistry, Federal State Unitary Enterprise,
Moscow, Russia
e-mail: yakunin@cc.nifhi.ac.ru

*Abstract*—In frameworks of the scaling theory of phase transitions and critical phenomena the quantitative dependence of macroscopic properties on nanostructural parameters in a polymeric material is revealed. The draw ratios at neck and at break are referred to the macroscopic properties. The structure is characterized by an average thickness of amorphous layers in isotropic melt-crystallized linear high density polyethylene which is chosen as an example. The square of the neck draw ratio is equal to the product of the square of the draw ratio at break and the chain ends collision probability. This probability in its turn is proportional to the average thickness of amorphous layers in the isotropic material. The neck draw ratio is a parameter of order. Polymers with flexible chains are solutions in solid state as well as in melt, the interacting ends of a marked chain serving as a solvent. At critical polymerization degree all the phases are identical. This research is an important contribution to the molecular theory of polymer liquids. It has been found that the melt viscosity vs the molecular weight of linear flexible-chain polymer follows the power law with the 3.4-exponent within the reptation model near the critical point. This is different from the value 3 expected for the melt of ring macromolecules. The rotation vibration precession motion of chain ends about the polymer melt flow direction were taken into account to find better agreement with the experiment.

*Keywords-critical phenomena; polymer melts; entanglements; phase transitions; critical exponents*

## I. Introduction

The quantitative relationship of structural and macroscopic characteristics (respectively, an average thickness of amorphous layers and draw ratios at neck and at break) is revealed in linear melt-crystallized (MC) polyethylene (PE). This acquires a great significance for modern chemical nanoengineering when it is necessary to understand clearly how to obtain functional nanostructures in order to prepare the material with sought properties by self-organization and self-assembly. The model is associated with such phenomena as drawing polymeric materials and dissipating the energy during the transition from isotropic to oriented state. The results obtained are in a good agreement with experimental data.

It has been revealed [1] that the neck draw ratio, i.e. the ratio of initial cross-section area to final, $\lambda_n$, in linear MC PE decreases with increasing the mean-weight molecular mass, $M_w$, as $\ln\lambda_n = A - \beta\ln M_w$ where $A$ is a positive constant and $\beta \approx 0.3$ is the critical exponent of the fluctuation theory of the second order phase transition (PT). It has been also shown [1] that the transformation to the oriented state is the first order PT forming 3 stages: swelling the polymer under applied stress, dissolving crystallites and crystallizing extended macromolecules. From these data one can suppose that no drawing will occur ($\lambda_n = 1$) if the polymerization degree (the number of chain monomers per one macromolecule, PD) is more than a some critical meaning, $N_{cr}$. Otherwise, the free energy of oriented state is less than the free energy of isotropic state [1] if $N < N_{cr}$.

The polymer melt viscosity dependence, $\eta(N)$, is commonly described in terms of the reptation model [2-4] suggested by P.-G. de Gennes in 1971. The reptation theory explains why the power law for the melt viscosity relation to the molecular weight of polymer can be observed. However, this is not fully supported by the experimental data where the exponent is equal to *3.3-3.4* or even is out of the range compared to the theoretical value of *3*. In order to achieve a better agreement with the experiment one may assume that the length of the tube created by entanglement chains along which a macromolecule can crawl making the reptation motions, can be subjected to fluctuations [3]. Wool [5] obtained a similar result using a different concept. Other attempts have been aimed at modifying the underlying equations [6, 7], they still continue [8, 9]. Similar techniques [8] are believed to be applicable to study more complex systems such as phase-separate polymer solutions, blends, block and graft copolymer mesophases and other fluids. However, in such approaches the number of monomers between neighbour entanglements along the chain, $N_e$, is introduced phenomenologically or ignored completely. Moreover, the recent Monte Carlo results [10] have shown that reptation motions prevail in melts where entropic trapping is absent in contrast to swollen gels [11]. At last, there are models in which an exponential increase of the reptation time with increasing PD can be observed if the polymer chains are long enough $N > N_e^3$ [12, 13]. Thus, the questions dealing with polymer melt viscosity and entanglements have been extensively discussed [8-10, 12-14]. This indicates that the problems still exist although the entanglement concentration has been already estimated [5, 15].

One of aims of the article is to construct a solution in such a way that would enable to achieve a good agreement with the well-known experimental data within the classical reptation model [2] without any additional simplifying assumptions.

The structure of the article is following. We will define the critical PD or the critical point. Then we will recall the main principles of the reptation theory [2-4] and try to understand what ways can be used to modify the theory. We will demonstrate that the model [16] based on fluctuations of the

This work is financially supported by RFBR, grants 11-03-00669a.
Dedicated to the memory of Dr. V. A. Buchin from Moscow State University.

tube length may be not correct and suggest that a mechanical field be introduced to correct it. In that case, a chain can probably "swell" in a medium of other chains and the vector connecting the chain ends can rotate making a vibration precession motion about the direction of the polymer melt flow. Finally, we will obtain a good agreement with the experimental data using the modified reptation theory. Then we will compare correlation radii of concentration fluctuation above and below the melting temperature near the critical point, $N_{cr}$. They will be found approximately equal, this would further confirm our theory. Physical clarity of underlying concepts makes this approach very attractive in understanding and explaining the nature of energy dissipation in polymer systems.

## II. RESULTS AND DISCUSSION

### A. Calculation of Critical Polymerization Degree

It has been found [15] that the ordering parameter can be defined for the chain with excluded volume as $w^2 = (B/C) N^{-2\beta}$ and $w_{cr}^2 = (B/C)^2$ at $N = N_{cr}$ where $B \approx 0.2068$ is the constant for the Fisher probability density, $C \approx 2842.45$, $C/B \approx 13744.9$, the number of components of ordering field $n = 0$ [17] for the Wilson $\varepsilon$ – expansion [18]. If we determine

$$\lambda_n^2 = w^2/w_{cr}^2 = (C/B) N^{-2\beta} \quad (1)$$

and use the value of $\nu = 1-6^{-1/2}$, $\beta = 0.2998$ [1, 15] then $\ln N_{cr} \approx 15.89$. At $N = N_{cr}$ $\lambda_n = 1$ and the oriented state cannot be observed. These conclusions agree with the results of [1].

Receiving (1) it has been assumed [15] that the form of swelling coil is not spherical since there exists an fluctuation attraction between the chain ends owing to screening volume interactions. The mechanism of the interaction in polymer solutions [4] as well as in melts [19] was described.

Further we will use the Vidom-Kadanoff relation [19]

$$\gamma - 1 + (1 - \nu d) = -2\beta. \quad (2)$$

$\nu$ is the critical exponent of correlation radius, $d$ is the space dimension. Let us define the mean magnetic correlation for a magnetic system connecting with polymer one [19]

$$<M(r_0)M(r)> = (r_0/r)^{1+\zeta} \quad (3)$$

where $\zeta$ is the critical exponent of ordering field and the following relationship is true [19] $2\beta = \nu (d - 2 + \zeta)$. Taking into account that at critical point the correlation radius [19] $r \sim aN^\nu$ where $a$ is the diameter of monomer we see from (3) that

$$<M(r_0)M(r)> = (r_0/a)^{1+\zeta} N^{-2\beta}. \quad (4)$$

We connect the mean magnetic correlation with $\lambda_n^2$ (1).

Multiplying $\nu^{-1}\ln(r_0/r)$ by the Vidom-Kadanoff relation (2) we can find the exact thermodynamic equation

$$<M(r_0)M(r)> = g_E(r) P_C(r) \quad (5)$$

where $g_E(r) = (r_0/r)^{d-1/\nu}$ is proportional to the Edwards correlation function, $P_C(r) = (r/r_0)^{(\gamma-1)/\nu}$ is the des Cloiseuaxs probability of collision of chain ends [19].

Multiplying $\nu^{-1}\ln(n_0 r_0/100 a N^\nu)$ by the Vidom-Kadanoff relation (2) we can obtain the formulae

$$P_C(r) = (\lambda_n/\lambda_{br})^2 = l_a/l_a^{cr} \quad (6)$$

where $n_0 r_0/100a = (C/B)^{1/(1+\zeta)}$ (compare with (1)), $n_0/100$ is a constant, $l_a$ is the average thickness of amorphous layer in isotropic material, these layers containing the ends of a marked chain, and

$$\lambda_{br}^2 = (n_0/100)^{d-1/\nu} N_{es}^2 N^{1-\nu d} \quad (7)$$

$\lambda_{br}$ is the draw ratio at break. If $n_0 = 18.5$ then $r_0/a \approx 65631$ and $(18.5/100)^{d/2-1/2\nu} \approx 0.33$, $N_{es} = (r_0/a)^{d/2-1/2\nu} \approx 1430.4$ is another constant, the dependence $l_a \approx 0.5 \ l_K (100/n_0)^{(\gamma-1)/\nu} N^{\nu-1}$ agrees with experimental data [1], $l_K$ = 2nm is the Kuhn segment in linear PE, $l_a^{cr} \approx 0.5 \ l_K (r_0/a)^{(\gamma-1)/\nu} \approx 26.9$nm is the value of $l_a$ at $N = N_{cr}$. These results (7) are also in accordance with experimental data [20].

We see that at critical PD the correlation radius of concentration fluctuation $r_0/a \approx 6.6 \cdot 10^4$ is a macroscopic value. It should have approximately the same meaning above the melting temperature.

### B. Coefficient of rotation diffusion in polymer melts

A mechanism of motions associated with the chain ends has been considered by M. Doi [16]. He has suggested that on a short time scale the chain end moves around quickly within the distance $aN^{1/2}$. But on the terminal relaxation time scale the mean-square displacement of the centre of gravity for the coil should be equal approximately to the mean-square displacement for a separate Rouse segment [4]. The important conclusion follows this reasoning: no translational motions give additional contributions in viscosity. It should be taken into account rotation Brownian motions in order to use the reptation theory [2] without any changes.

Let us consider the mechanical field in which the frequency, $\omega$, of rotation vibration motion of chain about the polymer melt flow direction is much less than the reciprocal value of $\tau$: $\varphi = \omega\tau << 1$ where $\tau \propto N^3$ is the terminal relaxation time [2]. Then the arising phase difference, $\varphi$, can be assumed [21] to be connected with a resulting torque which tends to return the chain ends in their initial position. The scaling expression for the coefficient of rotation diffusion $D_{rot} \propto \tau^{-1}$ can be written as follows $D_{rot} \sim T\eta^{-1}g$ in order to satisfy the required condition for the true relaxation time, $\tau_{tr}$,

$$\tau_{tr} \sim \tau N^{1/2}. \quad (8)$$

Here, $T$ is the temperature expressed in energy units, $\eta$ is the viscosity, $g \sim a^{-2} r^{-1}$ is a pair correlation function [15] and we state that the torque has always to be applied with a certain (lever) length $r \sim aN^{1/2}$ where $r$ is the mean end-to-end distance of chain in melts [4, 19]. In more realistic case $r \sim aN^{1-\nu}$ [15], since, as we will see below, screening the volume interactions vanishes if the mechanical field is applied to a polymer melt. Consequently, the exponent sought $\approx 3.41$. Thus, the terminal time of relaxation (8) increases resulting in the hydrodynamic effect of velocity, $V$, of a ball-shaped body, rotating effectively in a high-molecular-weight liquid with the viscosity $\eta$, on the friction force $f$. A characteristic size of this body is $r$.

Let us write all the factors due to the rotation motions:

$$r \propto \tau_{tr}/\tau = N^{1-\nu}, \tag{9}$$

$$V \propto N^{-3\nu}, \tag{10}$$

$$f_1 aN/T \propto N^{4(1/2-\nu)} \tag{11}$$

where $\nu = 0.5$ for ideal chains and $> 0.5$ for excluded volume chains, $f_1$ is the friction force per one monomer [2-4, 19]. In order to observe the reptation motions we see that the following inequality should be held: $f_1 aN/T \leq 1$, i.e. the work against the friction force is less than the thermal energy, approximately. It is true for flexible enough chains ($f_1 aN \sim T$). Their conformations may be changed easy, and such chains will make these motions. Other assumptions lead to polymer glass with frozen conformations. Multiplying the both hands of this inequality by $aN$, we find

$$\tau_R V \leq aN \tag{12}$$

where $\tau_R \sim \zeta_1 a^2 N^2/T$ is the relaxation time of the first Rouse mode, $\zeta_1 = T/D_1$ is the friction coefficient of one monomer, $\tau_1 = a^2/D_1$ and $D_1$ are the time of relaxation and the diffusion coefficient typical for low-molecular-weight liquids, respectively. Then another form of the inequality (12) can be written as follows:

$$\xi^2 \geq R_g^2 \tag{13}$$

where $\xi^2 = Ta/f_1$ and the gyration radius $R_g \sim N^{2\nu-1/2}$ for the excluded volume chains [15]. For the ideal chains [2-4, 19] $r \sim R_g \sim R_0 \sim aN^{1/2}$.

It is the most important and interesting result of the present work. In reptational dynamics a characteristic spatial scale appears. Although it is connected with the friction force per one monomer, at least, it is of order of mean size of polymer coil.

We have seen from (13) that there exists a spatial $\xi \sim R_g$. This gives an opportunity to assume that if the reptation motions due to the action of a mechanical field in polymer melts take place indeed then i) screening the volume interactions [22] vanishes, ii) the fluctuation attraction of chain ends and other monomers [4, 19] remains, iii) the correlation radius can increase.

Note that although all the motions associated with the chain ends are fast at small scales of order of several atom radii, an inhibition of such motions of ends at large scales (of order of end-to-end distance) reduces the overall chain mobility. It is clear that the fluctuation attraction results in an inhibition of large-scale motion of chain ends as an average effect.

*C. Ring macromolecules*

It is impossible to compare the recent Monte Carlo research on linear and ring chains [23, 24] with our results for 2 reasons, although the investigations of ring macromolecules are continued expansively as seen from the literature [25-28]. The first one is reported by the authors. They are not capable of considering the case $N >> N_e$. The second one is the absence of hard criteria for decoupling the rotation and translation motions in their model for linear chains.

The self-diffusion coefficient for a melt of linear molecules can be estimated as follows: $D = R_0^2/\tau_{tr} \sim N^{(2\nu-1/2)2}/N^{4-\nu} \sim N^{-5(1-\nu)} \sim N^{-2.05}$ where we have assumed $R_0 \sim R_g$ and have used the expression for the gyration radius $R_g \sim N^{2\nu-1/2}$ obtained by the author [15] and used above. The reptation law $D \sim N^{-2}$ [19] can be expected for unknotted rings since there is no fluctuation attraction of the chain ends.

If $N < N_e$ then $D \sim (aN)^2/\tau_1 N^2 \sim D_1$ ($R_0 \sim r \sim aN$). Thus, for $N \sim N_e$ one can observe any exponent ranged from $-2.05$ to $0$. In this respect the estimation $N_e \approx 283$ for linear chains was obtained [15].

*D. Correlation radii near the melting temperature*

In order to compare the correlation radii of concentration fluctuation below (in solid state) and above (in melt) the melting point let us recall the definition of critical point [1, 29]. Below the melting temperature at $N = N_{cr}$ the neck draw ratio and the crystallinity tend to 1 and 0, respectively, $\ln N_{cr} \approx 15.89$ [1, 15, 29]. In [29] the ratio of the correlation radius of concentration fluctuation to the monomer diameter is equal to $r_0/a \approx 6.6 \cdot 10^4$. From (13) we can see that $\xi/a \geq R_g/a \approx 5.2 \cdot 10^4$ at $N = N_{cr}$, i.e. $r_0/a \approx \xi/a$. This confirms our considerations. At the critical PD the correlation radii of concentration fluctuation above and below the melting temperature are equal to each other approximately.

*E. Conclusion remarks*

The rotation vibration motion cannot be observed at macroscopic scales due to decreasing the field with increasing the distance. An antisymmetric stress seems can appear as a macromolecule is capable of making rotation vibration precession motions. At large scales it transforms to an average tensor which will be symmetric probably. We have also found that a non-equilibrium parameter of polymer melts such as the viscosity is indirectly connected with the equilibrium pair correlation function due to the finite length of macromolecules [15]. We can indeed neglect the fluctuation attraction of the chain ends, supposing it is little, for long enough chains, i.e. if $N \to \infty$ and the physical state of the melt is not changed as in [12, 13], but, in this case, at $N > N_{cr}$ the solid state and the melt are identical to each other.

Based on the molecular recognition of the ends of a marked chain due to their fluctuation attraction [4, 15, 19] the present theory enables not only to estimate critical exponents and to find the critical PD [15] but also to obtain the pair correlation function of concentration fluctuation (5) in order to connect the nanostructural characteristics with macroscopic polymer properties (6). The ends of a marked chain serve a solvent since in the solid state, on the one hand, the exponent $\gamma$ testifies to this fact, on the other hand, the viscous flow of melt weakens screening the monomer-monomer interactions and leads to enhanced effective attraction between the chain ends.

Both lamellar and fibrillar structures are dissipative since the crystal of infinite size has a minimum of free energy [30]. Thus, a methodology of research of non-equilibrium polymer systems has been elaborated.

ACKNOWLEDGMENT

Thanks are expressed to Professor A. R. Khokhlov and to Professor I. Ya. Erukhimovich from Moscow State University for useful discussions, to Professor S. N. Chvalun and to Professor A. N. Kraiko from Moscow Institute of Physics and Technology for helpful remarks, to Professor V. N. Pokrovskii from University of Malta for numerous comments and for the reprint [9], and also to Dr. A. V. Mironov and A. Titkov for technical assistance in preparing the manuscript.